\documentclass[aps,prl,twocolumn,showpacs,groupedaddress]{revtex4}
\usepackage{graphicx}
\usepackage{dcolumn}
\usepackage{bm}
\usepackage{ae}
\usepackage{aecompl}
\usepackage{amsmath}
\usepackage{amssymb}
\usepackage{epsfig}
\usepackage{multirow}

\begin{document}
\title{Irreducible finite-size effects in surface free energies from crystal-nucleation data.}
\author{T. Zykova-Timan$^{1,2}$,  C. Valeriani$^{3}$, E. Sanz$^{4}$, D. Frenkel$^{3}$  and E. Tosatti$^{2,5}$}
\affiliation{ $^{1}$ Dept. of Chemistry and Applied Biosciences, ETHZ,
6900 Lugano, Switzerland, $^{2}$SISSA and INFM/Democritos, Via Beirut 2-4, 34014 Trieste,
Italy, $^{3}$ FOM Institute for Atomic and
Molecular Physics, Kruislaan 407, 1098 SJ Amsterdam, The
Netherlands, $^{5}$ Dept. of Physics, Utrecht University,
Princetonplein 5, 3584CC Utrecht, The Netherlands, $^{5}$ The Abdus Salam ICTP, 34014 Trieste, Italy}
\date{\today}

\begin{abstract}
In this Letter we report a simulation study in which we compare
the solid-liquid interfacial free energy of NaCl at coexistence,
with the value that follows from the height of the homogeneous
nucleation barrier. We find that the two estimates differ by more
than 100\%. Similar, although smaller discrepancies are found for
crystals of hard-sphere colloids and of Lennard-Jones (``argon'')
particles. We consider a variety of possible causes for this
discrepancy and are forced to conclude that it is due to a
finite-size effect that cannot be corrected for by any simple
thermodynamic procedure. Importantly, we find that the surface
free energies that follow from real nucleation experiments should
be subject to large finite size effects. Taking this in to
account, we obtain quantitative agreement between the simulation
data and the surface free energy of NaCl that follows from
nucleation experiments.  Our finding suggests that most
published solid-liquid surface free energies derived from
nucleation experiments will have to be revised.
\end{abstract}

\maketitle The study of homogeneous crystal nucleation is of
interest because it provides information about the pathway by
which crystalline order emerges from the disordered parent phase.
However, such experiments are also of considerable practical
importance, as they are used to estimate the magnitude of the
solid-liquid interfacial free energy.  Classical Nucleation Theory
(CNT, see e.g.~\cite{Kelton}) provides the route by which
experimental nucleation rates are related to surface free
energies.  CNT relates the number of crystal nuclei that form per
second per cubic meter (denoted by $R$) to $\Delta G_{crit}$, the
height of the free-energy barrier that has to be crossed to
nucleate a crystal:$R= \kappa\;  e^{-\Delta G_{crit}/k_{B}T}$.
Here $\kappa$ is a kinetic prefactor, {\it T} is the absolute
temperature and $k_{B}$ is Boltzmann's constant. CNT predicts the
following expression for the height of the nucleation barrier:
$\Delta G_{crit} = c
\frac{\gamma_\mathrm{LS}^{3}}{\rho_{S}^{2}|\Delta \mu|^{2}}$,
where $\gamma_\mathrm{LS}$ is the liquid-solid surface free energy
per unit area,  $\Delta \mu$ is the difference in chemical
potential between the solid and the supercooled liquid, and
$\rho_{S}$ is the number density of the crystalline phase.   $c$
is a constant that depends on the shape of the cluster, e.g.
$c=16\pi/3$ for a spherical crystal nucleus. As the nucleation
rate depends exponentially on $\Delta G_{crit}$, the rate is a
very sensitive function of the surface free-energy density
$\gamma_{LS}$. A crucial assumption underlying CNT is that the
bulk and surface properties of a small crystal nucleus are the
same as those of a macroscopic crystal. However, it has been long
realized that this assumption is questionable, as a critical
crystal nucleus often contains only a few hundred molecules.
Indeed, in his review on crystal nucleation, Kelton writes:
``...while the precise meaning of [$\gamma_\mathrm{LS}$] is
uncertain, it constitutes a parameter that can be determined for
each element and profitably used to make predictions of the
nucleation behavior''. In other words: the $\gamma_\mathrm{LS}$
determined from nucleation experiments can only be used to predict
the outcome of other nucleation experiments, thus severely
limiting the predictive value of CNT.   More microscopic theories  
such as density-functional theory (DFT)~\cite{HarrowellOxtoby,lutsko},
the Cahn-Hilliard approach (CH)~\cite{granasy1},
or the phase-field formalism (PF)~\cite{granasy2} can,
and have been used to improve upon CNT. Yet, the question remains whether the widely-used CNT 
can be reformulated in such a way that it correctly describes the properties of small clusters yet, 
at te same time, reproduces the correct, macroscopic surface free energy.

Increasingly accurate simulation techniques allow us to probe both
the free energy of small nuclei and the surface free energies of
planar crystal-liquid interfaces. A case in point is the system
NaCl in contact with its melt. Ref.~\cite{ZykovaPRL05} reported
the surface free energy of a NaCl [100] interface in contact with
the coexisting liquid phase: $\gamma_\mathrm{LS}\approx 36\pm 6$
mJ m$^{-2}$. The effective surface free energy that follows from
the NaCl crystal-nucleation barrier at 800 K was reported in
ref.~\cite{Valsanfre}: $\gamma_\mathrm{LS}= 80\pm 1$ mJ m$^{-2}$
(assuming that the nucleus has a cubic shape). In addition,
nucleation experiments at 905K~\cite{NaClexp1} provide an
experimental estimate of $\gamma_\mathrm{LS}\approx$ 68 mJ
m$^{-2}$.

Another example of a large difference between $\gamma_\mathrm{LS}$
derived from the nucleation barrier and from coexistence data
comes from hard-sphere colloids: a comparison of simulations at
coexistence~\cite{davidchak05} and in the supersaturated
regime~\cite{AuerJCP 120 3015 2004} indicate that the value of
$\gamma_\mathrm{LS}$ estimated on the basis of the nucleation
barrier is some 30\% larger than the value for a planar interface
at coexistence.  A similar discrepancy (${\mathcal O}$(20\%))
exists between the surface free energy for the planar interface
and the crystal nucleus of the (truncated and (force-)shifted)
Lennard-Jones potential~\cite{gilmer2,david2,WoldeS4}. It is
clearly of considerable interest to understand the origin of the
discrepancy between the nucleation data and the results for
$\gamma_\mathrm{LS}$ at coexistence, as this might facilitate the
interpretation and analysis of experimental nucleation data.

In this Letter, we report a systematic study of the finite-size
effects in the  surface free energy of NaCl crystals in contact
with their melt. We chose this system because it shows the largest
discrepancy of all examples listed above. As in
refs.~\cite{ZykovaPRL05,Valsanfre,Zykova05} we use the Tosi-Fumi
rigid-ion-pair interaction potential~\cite{Tosi} to model the
inter-ionic interactions in NaCl. We note at the outset that it is
imprecise to speak of {\em the} surface free-energy density of a
small crystallite, as the value of $\gamma_\mathrm{LS}$ depends on
the choice of the dividing surface (equimolar dividing surface,
equi-enthalpy dividing surface, surface of tension etc. - see
ref.~\cite{WidomRowlinson}). For flat interfaces, the
corresponding surface free energies are all the same, but this is
not the case for strongly curved surfaces. The surface free energy
that enters into CNT is the one associated with the surface of
tension~\cite{mullins}. One property of the surface of tension is
that it is, to lowest order, independent of the choice of the
dividing surface. We use this property to determine
$\gamma_\mathrm{LS}$ associated with the surface of tension.  To
facilitate the comparison with the data of ref.~\cite{ZykovaPRL05}
that refer to a flat interface at coexistence, we deduce
$\gamma_\mathrm{LS}$ from the size-dependence of the free energy
of a small crystallite at coexistence. At coexistence, there is no
difference in chemical potential between the liquid and the bulk
solid, hence the excess free energy of a small crystallite is
entirely due to its surface.

\begin{figure}[h!]
\centering
\includegraphics[clip,width=0.25\textwidth,angle=-0]{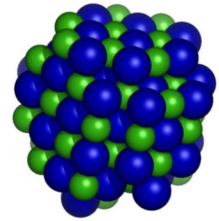}
\caption[a]{Average shape of a NaCl crystallite consisting of
140 solid-like particles. As the shape of such a small
cluster fluctuates, we obtain the average shape by superimposing a
large number of instantaneous configurations of the same mass.
fixing their center of mass, and the orientation of the crystal
axes. We average over all 48 symmetry-related orientations. The
surface is defined as the set of points where the average density
equals the average of the solid and liquid densities. Only the
crystalline  particles inside this surface are shown.
\label{fig:clustershape} }
\end{figure}

\begin{figure}[h!]
\begin{center}
\includegraphics[clip,width=0.4\textwidth,angle=-0]{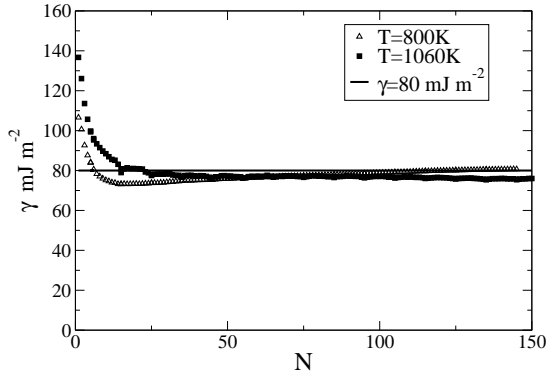}
\end{center}
\caption{Interfacial free energy as a function of $N$ computed at
the surface of tension, assuming a  cubic cluster  with $a$=0}
\label{gammaflat}
\end{figure}

All simulations were carried out at the coexistence temperature
$T_\mathrm{M}$=(1060$\pm$10)~K. $T_\mathrm{M}$ reported
in~\cite{Zykova05,Jamshed}. The melting temperature of Tosi-Fumi
NaCl is close to the experimental value:
$T_\mathrm{M}^{exp}$=1074~K~\cite{Janz}. As a first step, we
determine the dependence of the free energy of small NaCl
crystallites on the number of ions in the crystal. For this part
of the calculation, we make use of umbrella
sampling~\cite{umbrella} at constant  $N, P$ and $T$. These
simulations yield the excess free energy of the largest
crystalline cluster in the system as a function of  the number of
particles in that cluster. We use a geometrical criterion (see
ref.~\cite{Valsanfre}) to distinguish crystalline from liquid-like
particles. We can then deduce the surface free-energy density

using
\begin{equation}
\gamma_\mathrm{LS}(N) = \frac{\Delta G(N)}{C}  \ \rho_{s}^{2/3}
N^{-2/3}. \label{usedeq_a}
\end{equation}
where $N$ denotes the number of crystalline particles and $C$ is a
geometrical constant that has a value 6 for a cubic nucleus and
$(36\pi)^{1/3}$ for a sphere. Although there are strong
fluctuations in the shape of a small NaCl crystallite in contact
with its melt, its average shape is fairly cubic (see
figure~\ref{fig:clustershape}).
Of course, we need not assume {\em a priori}\ that the cluster is
cubic: we can use the average cluster shape from
figure~\ref{fig:clustershape} to perform a Wulff construction (see
e.g.~\cite{nozieres}) that yields the variation of the surface
free energy with orientation. Assuming that the surface free
energy of the $[100]$ equals the macroscopic value, we can the
compute  the average $\gamma_\mathrm{LS}$ of the cluster. Leaving
apart the question whether a Wulff construction is at all
meaningful for clusters containing ${\mathcal O}(10^2)$ particles,
we note that this procedure yields $\langle \gamma_\mathrm{LS}
\rangle\approx$ 40 mJ m$^{-2}$, which is within 10\% of the value
expected for a perfect cube. In what follows, we will therefore
assume that small NaCl crystals have the same cubic morphology as
macroscopic crystals.
\begin{figure}[h!]
\centering
\includegraphics[clip,width=0.4\textwidth,angle=-0]{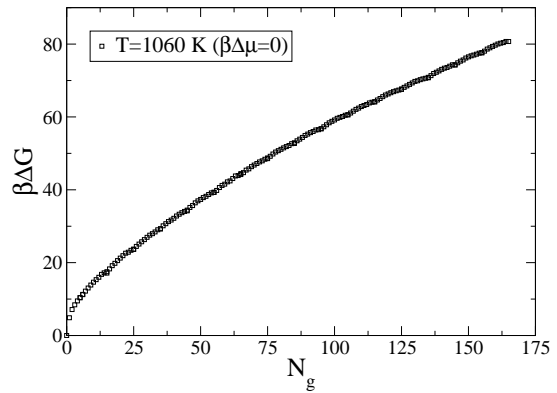}
\caption{Free energy barriers ($\beta \Delta G$) as a function of
$N_g$ at coexistence  $T_\mathrm{M}$. The error bars in $\beta
\Delta G$ are comparable to the size of the
symbols.\label{barriere}}
\end{figure}
From Figure~\ref{barriere}, we cannot yet deduce the surface free
energy because there is no {\em a priori} reason to assume that
the surface of this geometrical cluster has any thermodynamic
meaning. We know, however, that in the thermodynamic limit, the
ratio of $N_g$ to $N$, the ``thermodynamic''  number of atoms in
the crystal, should approach 1. We therefore make the ansatz:
$N=(N_g^{1/3}+a)^3$, where $a$ is an adjustable parameter that
remains to be determined. To find the number of atoms within the
surface of tension, we choose a value of $a$ that minimizes the
variation of $\gamma_\mathrm{LS}$ with the size of the cluster.
This analysis leads to a value of $a\approx0$.
Figure~(\ref{gammaflat}) shows that, over range of cluster sizes
studied, the resulting value of $\gamma_\mathrm{LS}$ is indeed
almost independent of $N$ for all but the smallest clusters. More
interestingly, we find that the resulting value of
$\gamma_\mathrm{LS}$ is very close to the value
$\gamma_\mathrm{LS}\approx 80$mJ m$^{-2}$ that follows from the
analysis of the nucleation barrier at 800K~\cite{Valsanfre} (see
figure~\ref{gammaflat}). Moreover, a similar analysis at 800 K,
leads to the same estimate of $\gamma_\mathrm{LS}$.
The internal consistency between the values of
$\gamma_\mathrm{LS}$ derived from the nucleation barrier and from
the surface of tension would be encouraging, were it not for the
fact that it does nothing to resolve the discrepancy with the
value of 36 mJ m$^{-2}$ found for a flat interface. Choosing
another conventional dividing surface (e.g. the equimolar or the
equi-enthalpy surface)~\cite{noteDivSurf} only makes matters
worse: in both cases we find a negative value of $a$ that results
in an even larger value of $\gamma_\mathrm{LS}$ that is, moreover,
strongly cluster-size dependent.  Hence, we conclude that the
discrepancy between the properties of a small crystal and a flat
interface cannot be corrected for by choosing a better definition
of the location of the solid-liquid interface.

The above discussion suggests that the conventional version of CNT
cannot account for the observed discrepancy between the surface
free energy of a flat interface and that of a small crystallite.
However, even within a thermodynamic approach, one can introduce
corrections to classical nucleation theory that would change the
apparent value of the surface free energy. One such correction
takes into account that the crystal nucleus is compressible and
that the surface free energy depends on the density of the
crystal. To estimate the magnitude of this effect, we extend the
analysis of Mullins~\cite{mullins} to obtain:
\begin{displaymath}\label{StressCompress2}
r\approx
\frac{(\gamma_\mathrm{LS}(\rho_s)/\gamma_\mathrm{LS}(\rho_s(0))^3}{
\left(1-\Delta\mu\rho_s/(2B) -\frac{1}{2} B
\epsilon^2/(\rho_s\Delta\mu)\right)^2}
\end{displaymath}
where $r$ is the ratio between the barrier height in the case of
compressible nuclei (with density $\rho_s$), compared to that for
incompressible clusters (with density $\rho_s(0)$). $B$ denotes
the bulk modulus of the crystal and $\epsilon$ the elastic strain,
compared to that of a solid at the same chemical potential as that
of the parent liquid. From our simulations, we find that the
density at the center of the crystal nucleus is some 6\% lower
than the reference value.  Even with this rather extreme estimate
of the strain in the nucleus, we find a compressibility-correction
to the apparent value of $\gamma_\mathrm{LS}$ that is no more than
10\%. Hence, we conclude that compressibility effects cannot
account for the observed discrepancy.

Thus far, we have not considered the effect of edges and vertices
on the surface free energy of a small cluster. This effect is
certainly non-negligible. If, for instance, we consider a cubic
NaCl crystal in vacuum at $T=0K$, both the line energy of the
edges and the vertex energy of the corners can be determined
directly. The energy of an NaCl cube can be written as:
\begin{equation}
e = e_B \ell^3 + 6 e_{S} \ell^{2} + 12 e_E \ell  + 8 e_C
\label{fitEC0KP}
\end{equation}
where $e$ is the total internal energy per particle, $e_B$, the
energy per particle in a bulk crystal,  $e_{S}$  the energy of a
particle belonging to the surface, $e_{E}$ the energy of a
particle belonging to an edge, $e_{C}$ the energy of a particle
belonging to a corner of the cube, and $\ell$ is the number of
atoms per edge. Computing this energy for a crystal of 64, 216 and
512 atoms~\cite{espresso}, we find that $e_E/e_S=0.22$ and
$e_C/e_s=1.2$. The effect of these edge and vertex contributions
is to increase the apparent surface energy by 13\% for a crystal
of 216 particles. Of course, these numbers do not apply to a hot
NaCl crystal in contact with its melt and it is not at even
obvious how to define the various terms in that case, as not only
the magnitude but even the sign of $e_E$ and $e_C$ depend on the
precise choice of the dividing surface. This means that, within
the macroscopic framework imposed by CNT, we cannot reliably
estimate the edge and corner contributions to the surface free
energy.

We are therefore forced to conclude that the large apparent value
of $\gamma_\mathrm{LS}$ of small crystallites  is due to a
finite-size effect that is not easily accounted for by within a
``thermodynamic''  theory. Rather, the free energy of small
clusters must be computed using a molecular approach, either
theoretically (as in DFT~\cite{HarrowellOxtoby,lutsko},CH~\cite{granasy1},
or PF~\cite{granasy2}) or numerically,
as illustrated in the present work. In the present paper, and in
ref.~\cite{Valsanfre}, we computed the free energy of relatively
small clusters (up to $N$=200). However, under the experimental
conditions for crystal nucleation of NaCl (T=905K), the critical
nucleus is expected to contain
 ${\mathcal O}(6 \times 10^2)$ particles.
Calculations for larger clusters would be feasible, but expensive.
We therefore make the We therefore make the Tolman ``ansatz''
that the leading correction to surface free energy is proportional
to $1/R_c$, where $R_c$ is the radius of the critical nucleus. As
$1/R_c\sim \Delta\mu$, we assume that the variation in
$\gamma_\mathrm{LS}$ is of the form:
$\gamma_\mathrm{LS}(\Delta\mu)$=
$\gamma_\mathrm{LS}^\mathrm{coex}+ b\Delta\mu$. We can determine
$b$ from the simulation data of refs~\cite{ZykovaPRL05,Valsanfre}.
Inserting the value $\Delta\mu=0.3kT$ at T=905K, we predict that
under the condition of the nucleation experiments of
ref.~\cite{NaClexp1}, the effective value of $\gamma_\mathrm{LS}$
should be 67 mJ m$^{-2}$, in almost embarrassing agreement with
the experimental data ($\gamma_\mathrm{LS}$=68 mJ m$^{-2}$).
Although this good agreement is almost certainly fortuitous, it
does support our conjecture that the surface free energies
measured in nucleation experiments are subject to very large
finite size corrections (in this case: more than 80\%). If we take
this strong $\Delta\mu$-dependence of $\gamma_\mathrm{LS}$
seriously, it would mean that for strongly faceted crystals
(although not for NaCl), the nucleation barrier could start to
rise again  at large super-saturations. This should be
experimentally observable, as it would lead to an increase in the
final crystallite size in fully crystallized samples~\cite{Shi}.
Interestingly -- but we do not know if it is really relevant --
the final crystallite size in hard-sphere crystallization suddenly
grows as the concentration is increased beyond a volume fraction
of 58\%. If the barrier is a monotonically decreasing function of
the volume fraction, this should not happen.

In summary, our study of the  free energy of NaCl crystallites
indicates that the surface free energy is subject to large finite
size corrections that cannot be accounted for within a
thermodynamic theory.  Based on the small number of examples where
the relevant simulation data are available (NaCl, Lennard Jones,
hard spheres), we speculate that the finite size effects are most
pronounced for strongly faceted crystals, such as NaCl.  The
present results support the suggestion by Kelton that the large
number of published surface free energies that are based on
nucleation data are of little use to predict macroscopic surface
free energies. We stress that, in addition to nucleation studies, there are
other, more reliable, experimental routes to determine solid-liquid  surface free energies. An example is 
the grain-boundary groove method~\cite{groove}. However, such experiments are challenging,
especially for materials that have anisotropic surface free energies.
Our work highlights the fact that, if nucleation studies are used to estimate solid-liquid surface free energies, 
the analysis cannot be based on CNT but must make us of one of the more accurate, microscopic  theories for crystal nucleation that properly
account for the fact that crystal nuclei are far from macroscopic.

We gratefully acknowledge the assistance of Axel
Arnold~\cite{espresso}. Work in SISSA/ICTP/Democritos was
sponsored by {\em PRIN-COFIN 2006022847}, as well as by INFM,
``Iniziativa trasversale calcolo parallelo''. The work of the
FOM Institute is part of the research program of FOM and is made
possible by financial support from the Netherlands organization
for Scientific Research (NWO), partly through grant 047.016.001.
An NCF grant for computer time is gratefully acknowledged.

\begin{thebibliography}{20}
\bibitem{noteDivSurf}
The equimolar dividing surface is defined such there are no
surface excess particles. This means that the total number of
particles in the system ($N_{tot}$) is equal to the sum of the
number of solid and liquid particles, $N_{tot}=N_S+N_L$ and
 the total volume in the system, $\langle V \rangle= \langle v_L \rangle N_L + \langle v_S
\rangle N_S$, where $v_L$ and $v_S$ are the (known) volumes per
particle, in the  liquid and solid phase, respectively. In a
simulation, we measure $\langle V \rangle$ as a function of $N_g$.
Assuming, as before, that $N_S=(N_g^{1/3}+a)^3$, we adjust $a$
such that we obtain the best linear relation between $\langle V
\rangle$ and $N_S$. The analysis of the equi-enthalpy surface
proceeds in a similar way.
\end{thebibliography}

\end{document}